\documentstyle[sprocl,twoside]{article}

\def\be{\begin{equation}}
\def\ee{\end{equation}}
\def\bea{\begin{eqnarray}}
\def\eea{\end{eqnarray}}

\newcommand{\AAA}{{\cal A}}

\def\esp#1{e^{#1}}
 \def\bra#1{\langle #1 |}
\def\ket#1{| #1\rangle }
\def\Zint{{Z \kern -.45 em Z}}
\def\complex{{\kern .1em {\raise .47ex \hbox
{$\scriptscriptstyle |$}} \kern -.4em {\rm C}}}
\def\real{{\vrule height 1.6ex width 0.05em depth 0ex
\kern -0.06em {\rm R}}}
\newcommand{\dd}{\displaystyle}
\newcommand{\nn}{\nonumber}
\font\fortssbx=cmssbx10 scaled \magstep2

\begin{document}
$\vcenter{\hbox{\fortssbx University of Florence}}$
%
\hfill
$\vcenter{\hbox{\bf DFF-379/12/01}}$ \vskip2.0cm \sloppy

\title{A THEORY OF ALGEBRAIC INTEGRATION}

\author{R. CASALBUONI}

\address{Department of Physics, University of Florence,\\ Via G. Sansone 1,
50019 Sesto Fiorentino (FI),  Italy}



\maketitle\abstracts{In this paper we extend the idea of
integration to generic algebras. In particular we concentrate
over a class of algebras, that we will call self-conjugated,
having the property of possessing equivalent right and left
multiplication algebras. In this case it is always possible to
define an integral sharing many of the properties of the usual
integral. For instance, if the algebra has a continuous group of
automorphisms, the corresponding derivations are such that the
usual formula of integration by parts holds. We discuss also how
to integrate over subalgebras. Many examples are discussed,
starting with Grassmann algebras, where we recover the usual
Berezin's rule. The paraGrassmann algebras are also considered,
as well as the algebra of matrices. Since Grassmann and
paraGrassmann algebras can be represented by matrices we show
also that their integrals can be seen in terms of traces over the
corresponding matrices. An interesting application is to the case
of group algebras where we show that our definition of integral
is equivalent to a sum over the unitary irreducuble
representations of the group. We show also some example of
integration over non self-conjugated algebras (the bosonic and
the $q$-bosonic oscillators), and over non-associative algebras
(the octonions).}

\newpage

\tableofcontents

\newpage
\section{Introduction}
\subsection{Motivations}
The very  idea of supersymmetry leads to the possibility of
extending ordinary classical mechanics to more general cases in
which ordinary configuration variables live together with
Grassmann variables. More recently the idea of extending
classical mechanics to more general situations has been further
emphasized with the introduction of quantum groups,
non-commutative geometry, etc. In order to quantize these general
theories, one can try two ways: i) the canonical formalism, ii)
the path-integral quantization. In refs.\cite{berezin,casalbuoni} classical theories involving Grassmann
variables were quantized by using the canonical formalism. But in
this case, also the second possibility can be easily realized by
using the Berezin's rule for integrating over a Grassmann
algebra.\cite{berezin2} It would be desirable to have a  way to perform
the quantization of theories defined in a general algebraic
setting. In this paper we will make a first step toward this
construction, that is we will give general rules allowing the
possibility of integrating over a given algebra. Given these
rules, the next step would be the definition of the
path-integral. In order to define the integration rules we will
need some guiding principle. So let us start by reviewing how the
integration over Grassmann variables comes about. The standard
argument for the Berezin's rule is translational invariance. In
fact, this guarantees the validity of the quantum action
principle. However, this requirement seems to be too technical
and we would rather prefer to rely on some more  physical
argument, as the one which is automatically satisfied by the path
integral representation of an amplitude, that is the combination
law for  probability amplitudes. This is a simple consequence of
the factorization properties of the functional measure and of the
additivity of the action. In turn, these properties follow in a
direct way from the very construction of the path integral
starting from the ordinary quantum mechanics. We recall that the
construction consists in the computation of the matrix element
$\langle q_f,t_f|q_i,t_i\rangle$, ($t_i<t_f$) by inserting the
completeness relation \be \int\; dq\;|q,t\rangle\langle q,t|=1
\label{completezza} \ee inside the matrix element at the
intermediate times $t_a$ ($t_i<t_a<t_f$, $a=1,\cdots,N$), and
taking the limit $N\to\infty$ (for sake of simplicity we consider
here the quantum mechanical case of a single degree of freedom).
The relevant information leading to the composition law is
nothing but the completeness relation (\ref{completezza}).
Therefore we will assume the completeness as the basic principle
to use in order to define the integration rules
 over a generic algebra. In
this paper we will limit our task to the construction of the
integration rules, and we will not do any attempt to construct the
functional integral in the general case. The extension of the
relation (\ref{completezza}) to a configuration space different
from the usual one is far from being trivial. However, we can use
an approach that has been largely used in the study of
non-commutative geometry\,\cite{connes} and  of quantum
groups.\cite{drinfeld} The approach starts from the observation that in
the normal case one can reconstruct a space from the algebra of
its functions . Giving this fact, one lifts all the necessary
properties in the function space and avoids to work on the space
itself. In this way can deal with cases in which no concrete
realization of the space itself exists. We will show  how to
extend the relation (\ref{completezza}) to the algebra of
functions. In Section 2 we will generalize these considerations
to the case of an arbitrary algebra. In Section 3 we will discuss
numerous examples of our procedure. Other examples will be given
in Sections 4 and 5. The approach to the integration of Grassmann
algebras starting from the requirement of completeness, which
inspired the present work, was discussed long ago by Martin.\cite{martin}

\subsection{The algebra of functions}

Let us consider a quantum dynamical system and an operator having
a complete set of eigenfunctions. For instance one can consider a
one-dimensional free particle. The hamiltonian eigenfunctions are
\be \psi_k(x)=\frac 1 {\sqrt{2\pi}}\exp{(-ikx)} \ee Or we can
consider the orbital angular momentum, in which case the
eigenfunctions are the spherical harmonics $Y_\ell^m(\Omega)$. In
general the eigenfunctions satisfy orthogonality relations \be
\int\psi_n^*(x)\psi_m(x)\;dx=\delta_{nm} \label{ortogonalita} \ee
(we will not distinguish here between discrete and continuum
spectrum). However $\psi_n(x)$ is nothing but the representative
in the $\langle x|$ basis of the eigenkets $|n\rangle$ of the
hamiltonian \be \psi_n(x)=\langle x| n\rangle \ee Therefore the
eq. (\ref{ortogonalita}) reads \be \int\langle n|x\rangle\langle
x|m\rangle\;dx=\delta_{nm} \ee which is equivalent to say that
the $|x\rangle$ states form a complete set and that $|n\rangle$
and $|m\rangle$ are orthogonal. But this means that we can
implement the completeness in the $|x\rangle$ space by using the
orthogonality relation obeyed by the eigenfunctions defined over
this space. On the other side, given this equation, and the
completeness relation for the set $\{\ket{\psi_n}\}$, we can
reconstruct the completeness in the original space $\real^1$,
that is the integration over the line. Now, we can translate the
completeness of the set $\{\ket{\psi_n}\}$, in the following two
statements
\begin{enumerate}
\item The set of functions $\{\psi_n(x)\}$ span a vector space.
\item The product $\psi_n(x)\psi_m(x)$ can be expressed as a linear
combination of the functions $\psi_n(x)$, since the set
$\{\psi_n(x)\}$ is complete.
\end{enumerate}
All this amounts to say that the set $\{\psi_n(x)\}$ is a basis
of an algebra. The product rules for the eigenfunctions are \be
\psi_m(x)\psi_n(x)=\sum_p c_{nmp}\psi_p(x) \label{1.6} \ee with
\be c_{nmp}=\int\psi_n(x)\psi_m(x)\psi_p^*(x)\;dx \label{cnmp} \ee
For instance, in the case of the free particle \be
c_{kk'k''}=\frac 1 {\sqrt{2\pi}}\delta(k+k'-k'') \ee Analogously,
for the angular momentum, one has the product
formula\,\cite{messiah} \bea Y_{\ell_1}^{m_1}(\Omega)
Y_{\ell_2}^{m_2}(\Omega)&=&
\sum_{L=|\ell_1-\ell_2|}^{\ell_1+\ell_2}\sum_{M=-L}^{+L}\left[
\frac{(2\ell_1+1)(2\ell_2+1)}{4\pi(2L+1)}\right]\nn\\
&\times&\langle\ell_1\ell_2 0 0| L0\rangle\langle \ell_1 \ell_2
m_1 m_2| LM\rangle Y_L^M(\Omega) \eea where $\langle j_1 j_1 m_1
m_2|JM\rangle$ are the Clebsch-Gordan coefficients. A set of
eigenfunctions can then be considered as a basis of the algebra
(\ref{1.6}), with structure constants given by (\ref{cnmp}). Any
function can be expanded in terms of the complete set
$\{\psi_n(x)\}$, and therefore it will be convenient, for the
future, to introduce the following ket made up in terms of
elements of the set $\{\psi_n(x)\}$ \be
|\psi\rangle=\left(\matrix{\psi_0(x)\cr\psi_1(x)\cr\cdots\cr\psi_n(x)
\cr\cdots\cr}\right) \ee A function $f(x)$ such that \be
f(x)=\sum_n a_n\psi_n(x) \ee
 can be represented as
\be f(x)=\langle a|\psi\rangle \ee where \be \langle
a|=\left(a_0,a_1,\cdots,a_n,\cdots\right) \ee To write the
orthogonality relation in terms of this new formalism it is
convenient to realize the complex conjugation as a linear
operation on the set $\{\psi_n(x)\}$. In fact, due to the
completeness, $\psi_n^*(x)$ itself can be expanded in terms of
$\psi_n(x)$ \be \psi_n^*(x)=\sum_nC_{nm}\psi_m(x),~~~ C_{nm}=\int
dx\,\psi_n^*(x)\psi^*_m(x) \ee or \be |\psi^*\rangle=C|\psi\rangle
\label{c_matrix} \ee Defining a bra as the transposed of the ket
$|\psi\rangle$ \be
\langle\psi|=(\psi_0(x),\psi_1(x),\cdots(x),\psi_n(x),\cdots) \ee
the orthogonality relation becomes \be
\int|\psi^*\rangle\langle\psi|\;dx=\int C|\psi\rangle\langle\psi|
\;dx=1 \label{newcompleteness} \ee Notice that  by taking the
complex conjugate of eq. (\ref{c_matrix}), we get \be C^*C=1 \ee
The relation (\ref{newcompleteness}) makes reference only to the
elements of the algebra of functions and it is the key element in
order to define the integration rules on the algebra. In fact, we
can now use the algebra product to reduce the  expression
(\ref{newcompleteness}) to a linear form \be
\delta_{nm}=\sum_\ell\int \psi_n(x)\psi_\ell(x)C_{\ell m}\; dx=
\sum_{\ell,p}c_{n\ell  p}C_{\ell m}\int \psi_p(x)\; dx \ee If the
set of equations \be \sum_p A_{nmp}\int\psi_p(x)\;
dx=\delta_{nm},~~~ A_{nmp}=\sum_\ell c_{n\ell  p}C_{\ell m} \ee
has a solution for $\int\psi_p(x)\; dx$, then we will be  able to
define the integration over all the algebra,  by linearity. We
will show in the following that indeed a solution exists for many
interesting cases.  For instance a solution always exists, if the
constant function is in the set $\{\psi_p(x)\}$. Let us  show
what we get for the free particle. The matrix $C$ is easily
obtained by noticing that \bea
\left(\frac 1{\sqrt{2\pi}}\exp(-ikx)\right)^*&=&\frac 1{\sqrt{2\pi}}\exp(ikx)\nn\\
&=&\int\;dk'\delta(k+k')\frac 1{\sqrt{2\pi}}\exp(-ik'x) \eea and
therefore \be C_{kk'}=\delta(k+k') \ee It follows \be
A_{kk'k''}=\int\;dq\;\delta(k'+q)\frac
1{\sqrt{2\pi}}\delta(q+k-k'')= \frac
1{\sqrt{2\pi}}\delta(k-k'-k'') \ee from which \be
\delta(k-k')=\int\;dk''\;\int\;A_{kk'k''}\psi_{k''}(x) dx=
\int\frac 1{2\pi}\exp(-i(k-k')x)dx \ee This example is almost
trivial, but it shows how, given the structure constants of the
algebra, the property of the exponential of being the Fourier
transform of the delta-function follows automatically from the
formalism. In fact, what we have really done it has been {\bf{ to
define the integration rules in the $x$ space}} by using only the
algebraic properties of the exponential. As a result, our
integration rules require that the integral of an exponential is
a delta-function. One can perform similar steps in the case of
the spherical harmonics, where the $C$ matrix is given by \be
C_{(\ell,m),(\ell',m')}=(-1)^m\delta_{\ell,\ell '}\delta_{m,-m'}
\ee and then using the constant function $Y_0^0=1/\sqrt{4\pi}$,
in the completeness relation.

 The procedure we have outlined here is the one  that we will
generalize in the next Section to arbitrary algebras. Before doing
that we will consider the possibility of a further
generalization. In the usual path-integral formalism sometimes
one makes use of the coherent states  instead of the position
operator eigenstates. In this case the basis in which one
considers the wave functions is a basis of eigenfunctions of a
non-hermitian operator \be \psi(z)=\bra\psi z\rangle \ee with \be
a\ket z=\ket z z \ee The wave functions of this type close an
algebra, as $\langle z^*|\psi^*\rangle$ do. But this time the two
types of eigenfunctions are not connected by any linear
operation. In fact, the completeness relation is defined on the
direct product of the two algebras \be \int\frac{dz^*dz}{2\pi
i}\exp(-z^*z)|z\rangle\langle z^*|=1 \ee Therefore, in similar
situations, we will not define the integration over the original
algebra, but rather on the algebra obtained by the tensor product
of the  algebra times a copy. The copy corresponds to the complex
conjugated functions of the previous example.
\section{Algebras}
\subsection{Self-conjugated algebras}
We recall here some of the concepts introduced in,\cite{integrale}
in order to define the integration rules over a
generic algebra. We start by considering an algebra $\AAA$ given
by $n+1$ basis elements $x_i$, with $i=0,1,\cdots n$ (we do not
exclude the possibility of $n\to\infty$, or of a continuous
index). We assume the multiplication rules \be x_i x_j=f_{ijk}x_k
\ee with the usual convention of sum over the repeated indices.
For the future manipulations it is convenient to organize the
basis elements $x_i$ of the algebra in a bra   \be \bra
x=\left(\matrix{x_0, & x_1, & \cdots & x_n}\right) \ee or in the
corresponding ket. Important tools for the study of a generic
algebra are the {\bf right and left multiplication algebras}. We
define the associated matrices by \be R_i|x\rangle=|x\rangle
x_i,~~~~\bra { x}L_i=x_i\bra{ x} \label{eigenequation} \ee For a
generic element $a=\sum_ia_ix_i$ of the algebra we have
$R_a=\sum_ia_iR_i$, and a similar equation for the left
multiplication. In the following we will use also \be L_i^T\ket
x=x_i\ket x \ee The matrix elements of $R_i$ and $L_i$ are
obtained from their definition \be
(R_i)_{jk}=f_{jik},~~~~(L_i)_{jk}=f_{ikj} \label{matrici} \ee The
algebra is completely characterized by the structure constants.
The matrices $R_i$ and $L_i$ are just a convenient  way of
encoding their properties. For instance, in the case of
associative algebras one has \be x_i(x_jx_k)=(x_ix_j)x_k \ee
implying the following relations (equivalent one with the other)
\be R_iR_j=f_{ijk}R_k,~~~L_iL_j=f_{ijk}L_k,~~~[R_i,L_j^T]=0
\label{associativity} \ee The first two say that  $R_i$ and $L_i$
are linear representations of the algebra, called the regular
representations. The third that the right and left
multiplications commute for associative algebras. In this paper
we will be interested in  algebras with identity, and such that
there exists a matrix $C$, satisfying \be L_i=CR_iC^{-1}
\label{self-conjugated} \ee We will call these algebras
self-conjugated. In the case of associative algebras, the
condition (\ref{self-conjugated}) says that the regular
representations (see eq. (\ref{associativity})) spanned by $L_i$
and $R_i$ are equivalent. Therefore, the non existence of the
matrix $C$ boils down two the possibility that the associative
algebra admits inequivalent regular representations. This
happens, for instance, in the case of the bosonic
algebra.\cite{integrale} In all the examples we will consider here, the
$C$ matrix turns out to be symmetric \be C^T=C \ee This condition
of symmetry  can be interpreted in  terms of the opposite algebra
$\AAA^D$, defined by \be x_i^D x_j^D=f_{jik}x_k^D \ee The left
and right multiplication in the dual algebra are related  to
those in $\AAA$ by \be R_i^D=L_i^T,~~~~L_i^D=R_i^T \ee Therefore,
in the associative case, the matrices $L_i^T$ are a representation
of the dual algebra \be L_i^TL_j^T\ket x=x_jx_i\ket
x=f_{jik}L_k^T\ket x \label{associativity2} \ee We see that the
property $C^T=C$ implies that the relation
(\ref{self-conjugated}) holds also for the right and left
multiplication in the opposite algebra \be L_i^D=CR_i^DC^{-1} \ee
In the case of associative algebras, the requirement of existence
of an identity is not a strong one, because we can always extend
the given algebra to another associative algebra with identity.\cite{schafer}
An extension of this type exists also for many
other algebras, but not for all. For instance, in the case of a
Lie algebra one cannot add an identity with respect to the Lie
product. For self-conjugated algebras, $L_i$ has an eigenket
given by \be L_i\ket{Cx}=CR_i\ket
x=\ket{Cx}x_i,~~~~\ket{Cx}=C\ket x \ee as it follows from
(\ref{self-conjugated}) and (\ref{eigenequation}). Then, as
explained in the Introduction, we define  the integration for a
self-conjugated algebra by the formula \be \int_{(x)}\ket
{Cx}\bra{x}=1 \label{2.21} \ee where 1 is the identity in the
space of the linear mappings on the algebra. In components the
previous definition means \be \int_{(x)}C_{ij}x_jx_k
=C_{ij}f_{jkp}\int_{(x)}x_p=\delta_{ik} \label{2.19} \ee This
equation is meaningful only if it is possible to invert it in
terms of $\int_{(x)} x_p$. This is indeed the case
 if $\AAA$ is an algebra with identity (say
$x_0=I$), \cite{integrale} because  by taking $k=0$ in eq.
(\ref{2.19}), we get \be \int_{(x)}x_j=(C^{-1})_{j0} \label{2.20}
\ee We see now the reason for requiring the condition
(\ref{self-conjugated}). In fact it ensures that the value
(\ref{2.20}) of the integral of an element of the basis of the
algebra gives the solution to the equation (\ref{2.19}). In fact
we have \be \int_{(x)}C_{ij}x_jx_k=C_{ij}f_{jkp}C^{-1}_{p0}=
(CR_k C^{-1})_{i0}=(L_k)_{i0}= f_{k0i}=\delta_{ik} \ee as it
follows from $x_kx_0=x_k$. Notice that if $C$ is symmetric we
can  write the integration also as \be \int_{(x)}\ket { x}\bra
{Cx}=1 \ee which is the form we would have obtained if we had
started with  the same assumptions but with the  transposed
version of  eq. (\ref{eigenequation}).
 We will define an
arbitrary function on the algebra by \be
f(x)=\sum_if_ix_i\equiv\langle x|f\rangle \ee and its conjugated
as \be f^*(x)=\sum_{ij}{\bar f}_iC_{ij} x_j=\langle f|Cx\rangle
\ee where \be \ket f=\left(\matrix{f_0 \cr f_1\cr
\cdot\cr\cdot\cr x_n}\right),~~~~ \bra f=\left(\matrix{{\bar f}_0
&{\bar f}_1 & \cdots &{\bar f}_n}\right) \ee and ${\bar f}_i$ is
the complex conjugated  of the coefficient $f_i$ belonging to the
field $\complex$. Then a scalar product on the algebra is given
by \be \langle f|g\rangle =\int_{(x)}\langle f|Cx\rangle \langle
x|g\rangle =\int_{(x)}f^*(x)g(x)= \sum_i{\bar f}_i g_i \ee

\subsection{Non self-conjugated algebras}

In order to generalize the case of coherent states seen at the
end of Section 1.2 we will consider now the case in which the $C$
matrix does not exist. For associative algebras this happens when
the left and right multiplications span inequivalent regular
representations. In this case, let us take an isomorphic copy of
${\cal A}$, say ${\cal A}^*$ \be x_i^*x_j^*=f_{ijk}x_k^* \ee and
\be R_i\ket {x^*}=\ket {x^*}
x_i^*,~~~~~\bra{x^*}L_i=x_i^*\bra{x_i^*} \ee with
$\ket{x^*}_i=x_i^*$. We then define the integration over the
direct product ${\cal A}\otimes{\cal A}^*$  as \be
\int_{(x,x^*)}|x^*\rangle\langle{x}|=1 \ee or \be
\int_{(x,x^*)}x_i^*x_j=\delta_{ij} \ee giving rise to the scalar
product \be \langle f|g\rangle=\int_{(x,x^*)}{\bar f}(x^*)  g(x)=
\sum_i{\bar f}_i g_i \ee

\subsection{Algebras with involution}

In  some case, as for the toroidal algebras,\cite{casalbuoni2}
the matrix $C$ turns out to define a mapping which is an
involution of the algebra. Let us consider the property of the
involution on a given algebra $\AAA$. An involution is a linear
mapping $^*:{\cal A}\to{\cal A}$, such that \be (x^{*})^*=x,~~~~~
(xy)^*=y^*x^*,~~~~x,y\in{\cal A} \label{definizioni} \ee
Furthermore, if the definition field of the algebra is
$\complex$, the involution acts as the complex-conjugation on the
field itself. Given a basis $\{x_i\}$ of the algebra, the
involution can be expressed in terms of a matrix $C$ such that \be
x_i^*=x_j C_{ji} \ee The eqs. (\ref{definizioni}) imply \be
(x_i^{*})^*=x_j^*C_{ji}^*=x_kC_{kj}C_{ji}^* \ee from which \be
 CC^*=1
 \label{quadrato}
\ee From the product property applied to the equality \be R_i\ket
x=\ket x x_i \ee we get \be (R_i\ket x)^*=\bra{x^*}
R_i^\dagger=\bra x CR_i^\dagger=(\ket x
x_i)^*=x_i^*\bra{x^*}=x_i^*\bra x C \ee and therefore \be \bra x
CR_i^\dagger C^{-1}=x_jC_{ji}\bra x= \bra x L_j C_{ji} \ee that is
\be CR_i^\dagger C^{-1}=L_jC_{ji} \ee or  \be CR_{x_i}^\dagger
C^{-1}=L_{x_i^*} \ee If $R_i$ and $L_i$ are $^*$-representations,
that is \be R_{x_i}^\dagger =R_{x_i^*}=R_{x_j}C_{ji} \ee we
obtain \be CR_{x_i}^\dagger C^{-1}=CR_{x_i^*}C^{-1}=L_{x_i^*}
\label{sopra} \ee Since the involution is non-singular, we get
\be CR_iC^{-1}=L_i \ee and comparing  with the adjoint of eq.
(\ref{sopra}), we see that $C$ is a unitary matrix which, from
eq. (\ref{quadrato}), implies $C^T=C$. Therefore we have the
theorem:
 \\\\ {\it  Given
an associative algebra with involution, if the right and left
multiplications are $^*$-representations, then the algebra is
self-conjugated.}\\\\ In this case our integration is a {\it
state}  in the Connes terminology.\cite{connes}

If the $C$ matrix is an involution we can write the integration as
\be
 \int_{(x)} \ket x\bra{x^*}=\int_{(x)} \ket {x^*} \bra x=1
 \ee

 \subsection{Derivations}

We will discuss now the derivations on  algebras with identity.
Recall that a derivation is a linear mapping on the algebra
satisfying \be D(ab)=(Da)b+a(Db),~~~~~a,b\in\AAA \label{2.47} \ee
We  define the action of $D$ on the basis elements in terms of
its representative matrix, $d$, \be Dx_i=d_{ij}x_j \ee If $D$ is
a derivation, then \be S=\exp(\alpha D) \ee is an automorphism of
the algebra. In fact, it is easily proved that \be \exp(\alpha
D)(ab)=(\exp(\alpha D)a)(\exp(\alpha D)b) \label{automorphism} \ee
On the contrary, if $S(\alpha)$ is an automorphism depending on
the continuous parameter $\alpha$, then from
(\ref{automorphism}), the following equation defines a derivation
\be D=\lim_{\alpha\to 0}\frac{S(\alpha)-1}\alpha \ee In our
formalism the automorphisms play a particular role. In fact, from
eq. (\ref{automorphism}) we get \be S(\alpha)(\ket x
x_i)=(S(\alpha)\ket x)(S(\alpha)x_i) \ee and therefore \be
R_i(S(\alpha)\ket x)=S(\alpha)(R_i\ket x)=S(\alpha)(\ket x
x_i)=(S(\alpha)\ket x)(S(\alpha)x_i) \ee meaning that
$S(\alpha)\ket x$ is an eigenvector of $R_i$ with eigenvalue
$S(\alpha)x_i$. This equation  shows that the basis
$x_i'=S(\alpha)x_i$ satisfies an algebra with  the same structure
constants as those of the  basis $x_i$. Therefore the matrices
$R_i$ and $L_i$ constructed in the two basis, and as a
consequence the $C$ matrix, are identical. In other words, our
formulation is invariant under automorphisms of the algebra (of
course this is not true for a generic change of basis). The
previous equation can be rewritten in terms of the matrix
$s(\alpha)$ of the automorphism $S(\alpha)$, as \be
R_i\left(s(\alpha)\ket x\right)=\left(s(\alpha)\ket x\right)
s_{ij}x_j=s_{ij}s(\alpha)R_j\ket x \ee or \be s(\alpha)^{-1}R_i
s(\alpha)=R_{S(\alpha)x} \label{auto} \ee If the algebra has an
identity element, $I$, (say $x_0=I$), then \be Dx_0=0 \label{di}
\ee and  therefore \be Dx_0=d_{0i}x_i=0 \Longrightarrow d_{0i}=0
\ee We will prove now some properties of the derivations. First
of all, from the basic defining equation (\ref{2.47}) we get \bea
R_id\ket x&=&R_i D\ket x=D(R_i\ket x=D(\ket x x_i) \nn\\&=&d\ket
x x_i+\ket x Dx_i= dR_i\ket x+ R_{Dx_i}\ket x \eea or \be
[R_i,d\,]=R_{Dx_i} \label{derivata} \ee which is nothing but the
infinitesimal version of eq. (\ref{auto}). From the integration
rules for a self-conjugated algebra with identity we get \be
\int_{(x)}Dx_i=d_{ij}\int_{(x)}x_j=d_{ij}(C^{-1})_{j0} \ee
Showing that in order that the derivation  $D$ satisfies the
integration by parts rule for any function, $f(x)$, on the algebra
\be \int_{(x)}D(f(x))=0 \label{integr_part} \ee the necessary and
sufficient condition is \be d_{ij}(C^{-1})_{j0}=0 \ee implying
that the $d$ matrix must be singular and have $(C^{-1})_{j0}$ as
a null eigenvector.

Next we show that, if a derivation satisfies the integration by
part formula (\ref{integr_part}), then the matrix of related
automorphism  $S(\alpha)=\exp(\alpha D)$ obeys the equation \be
Cs(\alpha)C^{-1}=s^{T-1}(\alpha) \label{invariance} \ee and it
leaves invariant the  integration. The converse of this theorem
is also true. Let us start assuming that $D$ satisfies eq.
(\ref{integr_part}), then \bea 0&=&\int_{(x)}D(C\ket x \bra x
)=\int_{(x)}Cd\ket x\bra x +\int_{(x)} C\ket x\bra {Dx}
\nn\\&=&CdC^{-1}+d^T \label{4.18} \eea that is \be CdC^{-1}=-d^T
\label{inv-infin} \ee The previous expression can be
exponentiated obtaining \be C\exp(\alpha d)C^{-1}=\exp(-\alpha
d^T) \ee from which the equation (\ref{invariance}) follows, for
$s(\alpha)=\exp(\alpha d)$. To show the invariance of the
integral, let us consider the following identity \be
1=\int_{(x)}s^{T-1}\ket {Cx}\bra{ x }s^{T}= \int_{(x)}Cs\ket
{x}\bra{ x }s^T =\int_{(x)}C\ket {Sx} \bra{ Sx}=\int_{(x)}C\ket
{x'} \bra{ x'} \label{5.22} \ee where $x'=Sx$, and we have used
eq. (\ref{invariance}). For any automorphism of the algebra we
have \be \int_{(x')}\ket {Cx'}\bra{ x'}=1 \label{5.23} \ee since
the numerical values of the matrices $R_i$ and $L_i$, and
consequently the $C$ matrix, are left invariant. Comparing eqs.
(\ref{5.22}) and (\ref{5.23}) we get \be \int_{(x')}=\int_{(x)}
\ee On the contrary, if the integral is invariant under an
automorphism of the algebra, the chain of equalities \be
1=\int_{(x')}\ket {Cx'}\bra{x'}=\int_{(x)}\ket
{Cx'}\bra{x'}=\int_{(x)}Cs\ket {x}\bra{x}s^T= CsC^{-1}s^T \ee
implies eq. (\ref{invariance}), together with its infinitesimal
version eq. (\ref{inv-infin}). From this (see the derivation in
(\ref{4.18})), we see that the following relation holds\be
0=\int_{(x)}D(C_{ij}x_jx_k) \ee and by taking $x_k=I$, \be
\int_{(x)}Dx_i=0 \ee for any basis element of
the algebra. Therefore we  have proven the following theorem:\\\\
{\it If a derivation $D$ satisfies the integration by part rule,
eq. (\ref{integr_part}), the integration is  invariant under the
related automorphism $\exp{(\alpha D)}$. On the contrary, if the
integration is invariant under a continuous automorphism,
$\exp{(\alpha D)}$, the related derivation, $D$, satisfies
(\ref{integr_part}).}\\\\ This theorem generalizes the classical
result about the Lebesgue integral relating the invariance under
translations of the measure and the integration by parts formula.


Next we will show that, always in the case of  an associative
self-conjugated algebra, $\AAA$, with identity, there exists  a
set of automorphisms  such that the measure of integration is
invariant. This is the of  the {\bf inner derivations}. In the
case of an associative algebra ${\cal A}$ with identity the set
coincides with the adjoint representation of Lie ${\cal A}$ (the
Lie algebra generated by $[a,b]=ab-ba$, for $a,~b\in\AAA$). That
is \be d= R_a-L_a^T\ee or \be Dx_i=x_i a- a x_i=-(adj
a)_{ij}x_j\ee

We can now proof the following theorem:\\\\{\it For an
associative self-conjugated algebra with identity, such that
$C^T=C$, the measure of integration is invariant  under the
automorphisms generated by the inner derivations, or,
equivalently, the inner derivations satisfy the rule of
integration by parts.}\\\\ In fact, this follows at once from eq.
(\ref{inv-infin}) \be
CdC^{-1}=C(R_a-L_a^T)C^{-1}=L_a-({C^{T}}^{-1}L_aC^T)^T=L_a-R_a^T=-d^T
\ee

\subsection{Integration over a subalgebra}

Let us start with a  self-conjugated algebra ${\cal A}$ with
generators $x_i$, $i=0,\cdots,n$. Let us further suppose that
${\cal A}$ has a self-conjugated sub-algebra ${\cal B}$ with
generators $y_\alpha$, with $\alpha=0,\cdots, m$, $m<n$. As a
vector space the algebra ${\cal A}$ can be decomposed as \be
{\cal A}={\cal B}\oplus {\cal C} \ee The vector space ${\cal C}$
is generated by vectors $v_a$, with $a=1,\cdots n-m$. Since
${\cal B}$ is a subalgebra we have multiplication rules \bea
y_\alpha y_\beta&=&f_{\alpha\beta\gamma}y_\gamma\nn\\
v_ay_\alpha&=&f_{a\alpha\beta}y_\beta+f_{a\alpha b}v_b\nn\\
v_a v_b&=&f_{abc}v_c+f_{ab\alpha}y_\alpha \eea By definition the
integration is defined both in ${\cal A}$ and in ${\cal B}$. Our
aim is to reconstruct the integration over ${\cal B}$ as an
integration over ${\cal A}$ with a convenient measure. To this
end, let us consider the matrix $S$ which realizes the change of
basis from $x_i$ to $(y_\alpha,v_a)$, that is \be
y_\alpha=S_{\alpha i}x_i,~~~~~v_a=S_{ai}x_i \ee This matrix is
invertible by hypothesis, and we can reconstruct the original
basis as \be x_i=(S^{-1})_{i\alpha}y_\alpha+(S^{-1})_{ia}v_a \ee
To reconstruct the integration over ${\cal B}$ in terms of an
integration over ${\cal A}$, we will construct a function on the
algebra \be P=p_i x_i \ee such that \be \int_{(\cal A)}v_a
P=0,~~~~~\int_{(\cal A)}y_\alpha P=\int_{(\cal B)}y_\alpha \ee
These are equivalent to require \be \int_{(\cal A)}{\cal
A}P=\int_{(\cal A)}{\cal B}P=\int_{(\cal B)}B \ee These are $n+1$
conditions over the $n+1$ unknown $p_i$. We will see immediately
that there is one and only one solution to the problem. In fact,
by using the matrix $S$ we can make more explicit the previous
equations by writing \be 0=\int_{(\cal A)}v_a
P=S_{ai}p_j\int_{(\cal A)}x_ix_j \ee Recalling that by definition
${\cal A}$ is \be \int_{(\cal A)}x_i x_j =(C_{\cal A}^{-1})_{ij}
\ee  we get \be 0=S_{ai}p_j(C_{\cal A}^{-1})_{ij} \ee and in
analogous way \be (SC_{\cal A}^{-1})_{\alpha j}p_j=\int_{(\cal
B)}y_\alpha \ee from which we obtain \be (SC_{\cal
A}^{-1})_{\alpha j}p_j=(C_{\cal B}^{-1})_{\alpha 0} \ee Since
both $S$ and $C$ are invertible, the problem has a unique
solution given by \be p_i=(C_{\cal A} S^{-1})_{i\alpha}(C_{\cal
B}^{-1})_{\alpha 0} \ee

\subsection{Change of variables}

Consider again a self-conjugated algebra and the following linear
change of variables \be x_i'=S_{ij}x_j \ee The integration rules
with respect to the new variables are \be \int_{x'}
x'_ix'_j=(C'^{-1})_{ij} \ee where $C'$ satisfies \be
L_i'=C'R_i'C'^{-1} \ee and the right and left multiplications in
the new basis are related to the ones in the old basis in the
following manner. From \be R_i'\ket {x'}=\ket{x'}x_i' \ee we get
\be R_i'S\ket x=S\ket x S_{ij}x_j=S_{ij}SR_j\ket x \ee or \be
R_i'=S_{ij}SR_jS^{-1} \ee and analogously \be
L_i'=S_{ij}S^{T-1}L_jS^T \ee In the new basis the $R$- and
$L$-representations are still equivalent ($L_i'=C'R_i'{C'}^{-1}$)
implying \be S^{T-1}L_iS^T=C'SR_iS^{-1}{C'}^{-1} \ee or \be
L_i=\left(S^TC'S\right)R_i\left(S^TC'S\right)^{-1} \ee Therefore
we must have ($L_i=CR_iC^{-1}$) \be C=S^TC'SA \label{2.115} \ee
with $A$ invertible and \be [R_i,A]=0 \ee We get \be
(C'^{-1})_{ij}=\int_{x'}x'_ix'_j=\int_{x'}S_{il}S_{jm}x_lx_m \ee
from which \be \int_{x'}x_lx_m=(S^{-1}C'^{-1}{S^T}^{-1})_{lm} \ee
and in particular \be \int_{x'}x_i=(S^{-1}C'^{-1}{S^T}^{-1})_{i0}
\ee The result can also be expressed in terms of the matrix $A$
defined in eq. (\ref{2.115}) \be A=S^{-1}C'^{-1}{S^{T}}^{-1}C \ee
obtaining \be \int_{x'}x_ix_j=(AC^{-1})_{ij} \ee
\section{Examples of Associative Self-Conjugated Algebras}
\subsection{The Grassmann algebra} We will discuss now the case of
the Grassmann algebra ${\cal G}_1$, with generators $1,\theta$,
such that $\theta^2=0$. The multiplication rules are \be
\theta^i\theta^j=\theta^{i+j},~~~ i,j,i+j=0, 1
\label{grassmannrules} \ee and zero otherwise. From the
multiplication rules we get the structure constants \be
f_{ijk}=\delta_{i+j,k},~~~~i,~j,~k=0,1 \ee from which  the
explicit expressions for the matrices $R_i$ and $L_i$ follow \bea
(R_0)_{ij}&=&f_{i0j}=\delta_{i,j}=\left(\matrix{ 1 & 0\cr
                                               0 & 1\cr}\right)\nn\\
(R_1)_{ij}&=&f_{i1j}=\delta_{i+1,j}=\left(\matrix{ 0 & 1\cr
                                               0 & 0\cr}\right)\nn\\
(L_0)_{ij}&=&f_{0ji}=\delta_{i,j}=\left(\matrix{ 1 & 0\cr
                                               0 & 1\cr}\right)\nn\\
(L_1)_{ij}&=&f_{1ji}=\delta_{i,j+1}=\left(\matrix{ 0 & 0\cr
                                               1 & 0\cr}\right)
\eea Notice that $R_1$ and $L_1$ are nothing but the ordinary
annihilation and creation Fermi operators with respect to the
vacuum state $\ket 0=(1,0)$.
 The $C$ matrix exists and it is given by
\be (C)_{ij}=\delta_{i+j,1}= \left(\matrix{ 0 & 1\cr
                  1 & 0\cr}\right)
\ee The eigenket of $R_i$ is \be
\ket\theta=\left(\matrix{1\cr\theta\cr}\right)
\label{g1ketandbra} \ee and the completeness reads \be
\int_{{\cal G}_1}\ket\theta\bra\theta= \int_{{\cal
G}_1}\left(\matrix{\theta & 0\cr
                                  1   & \theta\cr}\right)=
\left(\matrix{1 & 0\cr 0 & 1\cr}\right) \ee or \be \int_{{\cal
G}_1}\theta^i\theta^{1-j}=\delta_{i,j} \ee which means \be
\int_{{\cal G}_1}\, 1=0,~~~~\int_{{\cal G}_1}\,\theta=1
\label{3.8} \ee

The case of a Grassmann algebra ${\cal G}_n$, which consists of
$2^n$ elements obtained by $n$ anticommuting generators
$\theta_1,\theta_2,\cdots,\theta_n$, the identity, $1$, and by all
their products, can be treated in a very similar way. In fact,
this algebra can be obtained by taking a convenient tensor
product of $n$ Grassmann algebras ${\cal G}_1$, which means that
the  eigenvectors of the algebra of the left and right
multiplications are obtained by tensor product of the
eigenvectors of eq. (\ref{eigenequation}). The integration rules
extended by the tensor product give \be \int_{{\cal
G}_n}\theta_n\theta_{n-1}\cdots\theta_1=1 \ee and zero for all
the other cases, which is equivalent to require for each copy of
${\cal G}_1$ the equations (\ref{3.8}). It is worth to mention
the case of the Grassmann algebra ${\cal G}_2$ because it can be
obtained by tensor product of ${\cal G}_1$ times a copy ${\cal
G}_1^*$. Then we can apply our second method of getting the
integration rules and show that they lead to the same result with
a convenient interpretation of the measure. The algebra ${\cal
G}_2$ is generated by $\theta_1,\theta_2$. An involution of the
algebra is given by the mapping \be
^*:~~~~\theta_1\leftrightarrow \theta_2 \ee with the further rule
that by taking the $^*$ of a product one has to exchange the
order of the factors. It will be convenient to put
$\theta_1=\theta$, $\theta_2=\theta^*$. This allows us to consider
${\cal G}_2$ as ${\cal G}_1\otimes {\cal G}_1^*\equiv ({{\cal
G}_1,^*})$. Then the ket and bra eigenvectors of left and right
multiplication in ${\cal G}_1$ and ${\cal G}_1^*$ respectively are
given by \be \bra{{\theta}}=(1,\theta)~~~~~
\ket{\theta^*}=\left(\matrix{1\cr\theta^*\cr}\right) \ee with \be
R_i\ket{\theta^*}=\ket{\theta^*}{\theta^*}^i,~~~~~
\bra{{\theta}}L_i={\theta}^i\bra{{\theta}} \ee The completeness
relation  reads \be \int_{({{\cal
G}_1},^*)}\ket{\theta^*}\bra{{\theta}}= \int_{({{\cal
G}_1},^*)}\left(\matrix{1 & \theta\cr
                                  \theta^*   & \theta^*\theta\cr}\right)=
\left(\matrix{1 & 0\cr 0 & 1\cr}\right) \ee This implies \bea
\int_{({{\cal G}_1},^*)} 1&=&\int_{({{\cal
G}_1},^*)}\theta^*\theta=1\nn\\
\int_{({{\cal G}_1},^*)}\theta&=&\int_{({{\cal
G}_1},^*)}\theta^*=0 \eea These relations are equivalent to the
integration over ${\cal G}_2$ if we do the following
identification \be \int_{({{\cal G}_1},^*)}=\int_{{{\cal
G}_2}}\exp(\theta^*\theta) \ee The origin of this factor can be
traced back to the fact that we have \be \langle{{\theta}}|
{\theta}^*\rangle= 1+\theta\theta^*=\exp(-\theta^*\theta) \ee

\subsection{The Paragrassmann algebra}

 We will discuss now the case of a paragrassmann algebra of order $p$,
${\cal G}^p_1$, with generators $1$, and $\theta$, such that
$\theta^{p+1}=0$. The multiplication rules are defined by \be
\theta^i\theta^j=\theta^{i+j},~~~ i,j,i+j=0,\cdots, p
\label{paragrassmannrules} \ee and zero otherwise. From the
multiplication rules we get the structure constants \be
f_{ijk}=\delta_{i+j,k},~~~~i,~j,~k=0,1,\cdots,p \ee from which we
obtain the following expressions for the matrices $R_i$ and $L_i$:
\be (R_i)_{jk}=\delta_{i+j,k},~~~~(L_i)_{jk}=\delta_{i+k,j},~~~
i,j,k=0,1\cdots,p \label{3.19} \ee

The $C$ matrix exists and it is given by \be
(C)_{ij}=\delta_{i+j,p} \ee In fact \be (C R_i
C^{-1})_{lq}=\delta_{l+m,p}\delta_{i+m,n}\delta_{n+q,p}=
\delta_{i+p-l,p-q}=\delta_{i+q,l}=(L_i)_{lq} \ee The ket and the
bra eigenvectors of $L_i$ are  given by \be
C\ket\theta=\left(\matrix{\theta^p\cr\theta^{p-1}\cr\cdot\cr 1\cr}
\right),~~~~ \bra\theta=(1,\theta\cdots,\theta^p) \ee and the
completeness reads \be \int_{{\cal
G}^p_1}\theta^{p-i}\theta^{j}=\delta_{ij} \ee which means \be
\int_{{\cal G}^p_1}\, 1=\int_{{\cal G}^p_1}\,\theta= \int_{{\cal
G}^p_1}\,\theta^{p-1}=0 \ee \be \int_{{\cal G}^p_1}\,\theta^p=1
\ee in agreement with the results of ref.\cite{martin} (see
also\,\cite{isaev}).

\subsection{The algebra of matrices}

Since an associative algebra admits always a matrix
representation, it is interesting to consider the definition of
the integral over the algebra $\AAA_N$ of the $N\times N$
matrices. These can be expanded in the following general way \be
A=\sum_{n,m=1}^N e^{(nm)} a_{nm} \ee where $e^{(nm)}$ are $N^2$
matrices defined by \be
e^{(nm)}_{ij}=\delta_i^n\delta_j^m,~~~~i.j=1,\cdots,N
\label{e-matrices} \ee These special matrices satisfy the algebra
\be e^{(nm)}e^{(pq)}=\delta_{mp}e^{(nq)} \label{matrix-algebra}
\ee Therefore the structure constants of the algebra are given by
\be f_{(nm)(pq)(rs)}=\delta_{mp}\delta_{nr}\delta_{qs} \ee from
which \be
(R_{(pq)})_{(nm)(rs)}=\delta_{pm}\delta_{qs}\delta_{nr},~~~~
(L_{(pq)})_{(nm)(rs)}=\delta_{qr}\delta_{pn}\delta_{ms} \ee The
matrix $C$ can be found by requiring that $\ket {Cx}$ is an
eigenstate of $L_{pq}$, that is \be
(L_{(pq)})_{(nm)(rs)}[F(e)]^{(rs)}=[F(e)]^{(nm)}e^{(pq)} \ee where
\be F(e)^{(nm)}=C_{(nm)(rs)}e^{(rs)} \ee We get \be
[F(e)]^{(qm)}\delta_{pn}=[F(e)]^{(nm)}e^{(pq)} \ee By looking at
the eq. (\ref{matrix-algebra}), we see that this equation is
satisfied by \be [F(e)]^{(rs)}=e^{(sr)} \ee It follows \be
C_{(mn)(rs)}=\delta_{ms}\delta_{nr} \ee It is seen easily that
$C$ satisfies \be C^T=C^*=C, ~~~~C^2=1 \ee Therefore the matrix
algebra is a self-conjugated one. One easily checks that the
right multiplications satisfy  eq. (\ref{sopra}), and therefore
$C$ is an involution. More precisely, since \be
{e^{(mn)}}^*=C_{(mn)(pq)}e^{(pq)}=e^{(nm)} \ee the involution is
nothing but the hermitian conjugation \be A^*=A^\dagger, ~~A\in
\AAA_N \ee The integration rules give \be
(C^{-1})_{(rp)(qs)}=\delta_{rs}\delta_{pq}=
\int_{(e)}e^{(rp)}e^{(qs)}=\delta_{pq}\int_{(e)}e^{(rs)} \ee We
see that this is satisfied by \be \int_{(e)}e^{(rs)}=\delta_{rs}
\label{int-matrici} \ee This result can be obtained also using
directly eq. (\ref{2.20}), noticing that the identity of the
algebra is given by $I=\sum_n e^{(n,n)}$. Therefore \be
\int_{(e)}e^{(rs)}=\sum_n(C^{-1})_{(rs)(nn)}=\sum_n
\delta_{ns}\delta_{nr}= \delta_{rs} \ee and, for a generic matrix
\be \int_{(e)}A=\sum_{m,n=1}^Na_{nm}\int_{(e)}e^{(nm)}=Tr(A) \ee
Since the algebra of the matrices is associative, the inner
derivations are  given by \be D_B A=[A,B] \ee Therefore \be
\int_{(e)}D_B A=\int_{(e)}[A,B]=0 \ee and we see that the
integration by parts formula corresponds to the cyclic property
of the trace.

\subsection{The subalgebra ${\cal A}_{N-1}$}

Consider the algebra ${\cal A}_{N}$ of the $N\times N$ matrices,
and its subalgebra ${\cal A}_{N-1}$. We have the decomposition \be
{\cal A}_{N}={\cal A}_{N-1}\oplus{\cal C} \ee with \be {\cal
C}=\sum_{i=1}^{N-1}\left(e^{(i,N)}\oplus e^{(N,i)}\right)\oplus
e^{(N,N)} \ee and \be {\cal A}_{N-1}=\sum_{i,j=1}^{N-1}\oplus
e^{(i,j)} \ee Let us put \be P=\sum_{i,j=1}^N p_{ij}e^{(i,j)} \ee
then we require \be \int_{{\cal A}_N} {\cal C}P=0 \ee which
implies \be \int_{{\cal A}_N} e^{(i,N)} P= p_{Ni}=0 \ee and
analogously \be p_{iN}=p_{NN}=0 \ee The other condition \be
\int_{{\cal A}_N} {\cal A}_{{N-1}}P=\int_{{\cal A}_{N-1}} {\cal
A}_{{N-1}} \ee gives \be \int_{{\cal A}_N}e^{(i,j)}
P=p_{ji}=\delta_{ij} \ee Therefore \be P=\left(\matrix{ 1_{N-1} &
0 \cr 0 & 0 }\right) \ee where $1_{N-1}$ is the identity matrix
in $N-1$ dimensions. This result can be checked immediately by
computing the product $A_NP$ with $A$ a generic matrix of ${\cal
A_N}$ \be A_NP=\left(\matrix{ A_{N-1} & B \cr C & D }\right)
\left(\matrix{ 1_{N-1} & 0 \cr 0 & 0 }\right)= \left(\matrix{
A_{N-1} & 0 \cr C & 0 }\right) \ee implying \be \int_{{\cal
A}_N}A_NP=Tr(A_NP)=Tr(A_{N-1})=\int_{{\cal A}_{N-1}}A_{N-1} \ee

\subsection{Paragrassmann algebras as subalgebras of ${\cal A}_{p+1}$}

A paragrassmann algebra of order $p$ can be seen as a subalgebra
of the matrix algebra $A_{p+1}$. In fact, since this algebra is
associative it has a matrix representation (the regular one) in
terms the $(p+1)\times(p+1)$ right multiplication matrices, $R_i$
(see eq. (\ref{3.19}). These are given by \be
(R_i)_{jk}=\delta_{i+j,k} \ee Defining \be R_\theta\equiv R_1 \ee
we can write, in terms of the matrices defined in eq.
(\ref{e-matrices}) \be R_\theta=\sum_{i=1}^p e^{(i,i+1)} \ee and
\be R_\theta^k=\sum_{i=1}^{p+1-k}e^{(i,i+k)} \ee Therefore, the
most general function on  the paragrassmann algebra (as a
subalgebra of the matrices $(p+1)\times (p+1)$) is given by \be
f(R_\theta)=\sum_{i=1}^{p+1}a_iR_\theta^{p+1-i}=
\sum_{i=1}^{p+1}a_i\sum_{j=1} ^i e^{(j,p+1+j-i)} \ee In order to
construct the matrix $P$ defined in Section (2.5), let us
consider a generic matrix $B\in \AAA_{p+1}$. We can always
decompose it as (see later) \be B=f(R_\theta)+\tilde B
\label{decomposition} \ee In order to construct this
decomposition, let us consider the most general $(p+1)\times
(p+1)$ matrix. We can write \be B=\sum_{i,j=1}^{p+1}
b_{ij}e^{(ij)}=\sum_{i=1}^{p+1}\sum_{j=1}^p
b_{ij}e^{(ij)}+\sum_{i=1}^{p+1}b_{i,p+1}e^{(i,p+1)}
\label{B-matrix} \ee By adding and subtracting \be
\sum_{i=2}^{p+1}b_{i,p+1}\sum_{j=1}^{i-1}e^{(j,p+1+j-i)} \ee we
get the decomposition (\ref{decomposition}) with \be
f(R_\theta)=\sum_{i=1}^{p+1}b_{i,p+1}R_\theta^{p+1-i}
\label{effe-e} \ee and \be \tilde
B=\sum_{i=1}^{p+1}\sum_{j=1}^pb_{ij}e^{(ij)}-\sum_{i=2}^{p+1}
b_{i,p+1}\sum_{j=1}^{i-1}e^{(j,p+1+j-i)} \ee Now, we can check
that the matrix $P$ such that \be
\int_{\theta}f(\theta)=\int_{A_{p+1}}
BP=\int_{A_{p+1}}f(R_\theta)P \ee is given by \be P=e^{(p+1,1)}
\ee In fact, we have \be \tilde Be^{(p+1,1)}=0 \ee implying \be
Be^{(p+1,1)}=f(R_\theta)e^{(p+1,1)} \ee Furthermore \be
Tr[R_\theta^k
e^{(p+1,1)}]=\sum_{i=1}^{p+1-k}Tr[e^{(i,i+k)}e^{(p+1,1)}]=
Tr[e^{(p+1-k,1)}]=\delta_{p,k} \ee and therefore \be
\int_{\theta}\theta^k=Tr[R_\theta^k e^{(p+1,1)}]=\delta_{p,k} \ee
showing that $e^{(p+1,1)}$ is the matrix $P$ we were looking for.

We notice that the matrices $\tilde B$ and $f(R_\theta)$
appearing in the decomposition (\ref{decomposition}) can be
written more explicitly as \be \tilde B=\left(\matrix{ \tilde
b_{1,1} & \tilde b_{1,2} & \cdots & \tilde b_{1,p} & 0\cr
\cdot         &   \cdot       &   \cdot &    \cdot     &
\cdot\cr \cdot      &   \cdot        &       \cdot &       \cdot
&     \cdot\cr \tilde b_{p,1} & \tilde b_{p,2} & \cdots & \tilde
b_{p,p} & 0\cr \tilde b_{p+1,1} & \tilde b_{p+1,2} & \cdots &
\tilde b_{p+1,p} & 0\cr} \right) \ee and \be
f(R_\theta)=\left(\matrix{ a_{p+1} & a_p & a_{p-1} & \cdots  &a_2
&a_1\cr 0 & a_{p+1} & a_p & \cdots & a_3 & a_2\cr 0 & 0 & a_{p+1}
& \cdots & a_4 & a_3\cr \cdot & \cdot & \cdot & \cdot & \cdot &
\cdot\cr \cdot & \cdot & \cdot & \cdot & \cdot & \cdot\cr
 0 & 0 & 0 & \cdots & a_{p} & a_{p-1} \cr
 0 & 0 & 0 & \cdots & a_{p+1} & a_p \cr
 0 & 0 & 0 & \cdots & 0 &
a_{p+1}\cr}\right) \ee The $p\times (p+1)$ parameters appearing
in $\tilde B$ and the $p+1$ parameters in $f(R_\theta)$ can be
easily expressed in terms of the $(p+1)\times (p+1)$ parameters
defining the matrix $B$.

In the particular case of a Grassmann algebra we have \be
R_\theta=e^{(1,2)}=\left(\matrix{0 & 1\cr 0 &
0}\right)=\sigma_+,~~~ P=e^{(2,1)}=\left(\matrix{0 & 0\cr 1 &
0}\right)=\sigma_- \ee The decomposition in eq.
(\ref{decomposition}), for a $2\times 2$ matrix \be
B=a+b\sigma_3+c\sigma_++d\sigma_- \ee is given by \be \tilde
B=b(1+\sigma_3)+d\sigma_-,~~~~f(R_\theta)=f(\sigma_+)=
a-b+c\sigma_+ \ee and the integration is \be
\int_{(\theta)}f(\theta)=Tr[f(\sigma_+)\sigma_-] \ee from which
\be \int_{(\theta)} 1=Tr[\sigma_-]=0,~~~~ \int_{(\theta)}
\theta=Tr[\sigma_+\sigma_-]=1 \ee

\subsection{Projective group algebras}

Let us start defining a projective  group algebra. We consider an
arbitrary projective linear representation, $a\to x(a)$, $a\in
G$, $x(a)\in \AAA(G)$, of a given group $G$. The representation
$\AAA(G)$ defines in a natural way an associative algebra with
identity  (it is closed under multiplication and it defines a
generally complex vector space). This algebra will be
 denoted by $\AAA(G)$. The elements of the algebra
are given by the combinations \be \sum_{a\in G}f(a) x(a)
\label{groupalgebra} \ee For a group with an infinite number of
elements, there is no a unique definition of such an algebra. The
one defined in eq. (\ref{groupalgebra}) corresponds to consider a
formal linear combination of a finite number of elements of $G$.
This is very convenient since we will not be concerned here with
topological problems. Other definitions correspond to take
complex functions on $G$ such that \be \sum_{a\in G}|f(a)|<\infty
\ee Or, in the case of compact groups, the sum is defined in
terms of the Haar
 invariant measure. In the following we  will not need to
be more precise about this point. The basic product rule of the
algebra follows from the group property \be
x(a)x(b)=\esp{i\alpha(a,b)}x(ab) \label{algebra} \ee where
$\alpha(a,b)$ is called a cocycle. This is constrained, by the
requirement of associativity of the representation, to satisfy

\be \alpha(a,b)+\alpha(ab,c)=\alpha(b,c)+\alpha(a,bc)
\label{cocycle1} \ee Changing the element $x(a)$ of the algebra
by a phase factor $\esp{i\phi(a)}$, that is, defining \be
x'(a)=\esp{-i\phi(a)}x(a) \ee we get \be
x'(a)x'(b)=\esp{i(\alpha(a,b)-\phi(ab)+\phi(a)+\phi(b))}x'(ab) \ee
This is equivalent to change the cocycle to \be
\alpha'(a,b)=\alpha(a,b)-[\phi(ab)-\phi(a)-\phi(b)] \ee In
particular, if $\alpha(a,b)$ is of the form
$\phi(ab)-\phi(a)-\phi(b)$, it can be transformed to zero, and
therefore the corresponding projective representation is
isomorphic to a vector one. For this reason the combination \be
\alpha(a,b)=\phi(ab)-\phi(a)-\phi(b) \ee is called a trivial
cocycle. Let us now discuss some properties of the cocycles. We
start from the relation ($e$ is the identity element of $G$) \be
x(e)x(e)=\esp{i\alpha(e,e)}x(e) \ee By the transformation
$x'(e)=\esp{-i\alpha(e,e)}x(e)$, we get \be x'(e)x'(e)=x'(e) \ee
Therefore we can assume \be \alpha(e,e)=0 \ee Then, from \be
x(e)x(a)=\esp{i\alpha(e,a)}x(a) \ee multiplying by $x(e)$ to the
left, we get \be x(e)x(a)=\esp{i\alpha(e,a)}x(e)x(a) \ee implying
\be \alpha(e,a)=\alpha(a,e)=0 \ee where the second relation is
obtained in analogous way. Now, taking $c=b^{-1}$ in eq.
(\ref{cocycle1}), we get \be
\alpha(a,b)+\alpha(ab,b^{-1})=\alpha(b,b^{-1}) \label{cocycle2}
\ee Again, putting $a=b^{-1}$ \be
\alpha(b^{-1},b)=\alpha(b,b^{-1}) \ee We can go farther by
considering \be x(a)x(a^{-1})=\esp{i\alpha(a,a^{-1})}x(e) \ee and
defining \be x'(a)=\esp{-i\alpha(a,a^{-1})/2}x(a) \ee from which
\be
x'(a)x'(a^{-1})=\esp{-i\alpha(a,a^{-1})}x(a)x(a^{-1})=x(e)=x'(e)
\ee Therefore we can transform $\alpha(a,a^{-1})$ to zero without
changing the definition of $x(e)$, \be \alpha(a,a^{-1})=0
\ee As a consequence,  equation (\ref{cocycle2}) becomes \be
\alpha(a,b)+\alpha(ab,b^{-1})=0 \label{cocycle3} \ee We can get
another relation using $x(a^{-1})=x(a)^{-1}$ \bea
 x(a^{-1})x(b^{-1})&=&\esp{i\alpha(a^{-1},b^{-1})}x(a^{-1}b^{-1})=
 x(a)^{-1}x(b)^{-1}\nn\\&=&(x(b)x(a))^{-1}=
 \esp{-i\alpha(b,a)}x(a^{-1}b^{-1})
\eea from which \be \alpha(a^{-1},b^{-1})=-\alpha(b,a)
\label{3.132} \ee and together with eq. (\ref{cocycle3}) we get
\be \alpha(ab,b^{-1})=\alpha(b^{-1},a^{-1}) \label{cocycle4} \ee
The last two relations will be useful in the following. From the
product rule \be x(a)x(b)=\esp{i\alpha(a,b)}x(ab)=\sum_{c\in G}
f_{abc}x(c) \ee we get the structure constants of the algebra \be
f_{abc}=\delta_{ab,c}\esp{i\alpha(a,b)} \ee The delta function is
defined according to the nature of the sum over the group
elements.

To define the integration over $\AAA(G)$, we start as usual by
introducing a ket with elements given by $x(a)$, that is $\ket
{x}_a=x(a)$, and the corresponding transposed bra $\bra x$. From
the algebra product, we get immediately \be
(R(a))_{bc}=f_{bac}=\delta_{ba,c}\esp{i\alpha(b,a)},~~~~
(L(a))_{bc}=f_{acb}=\delta_{ac,b}\esp{i\alpha(a,c)} \ee We show
now that also these algebras are self-conjugated. Let us look for
eigenkets of $L(a)$ \be L(a)\ket {Cx}=\ket{Cx}x(a) \label{2.31}
\ee giving \be \delta_{ac,b}(Cx)_c\esp{i\alpha(a,c)}=
\esp{i\alpha(a,a^{-1}b)}(Cx)_{a^{-1}b}=(Cx)_b x(a) \ee By putting
\be (Cx)_a=k_ax(a^{-1}) \ee we obtain \be
k_{a^{-1}b}x(b^{-1}a)\esp{i\alpha(a,a^{-1}b)}=
k_b\esp{i\alpha(b^{-1},a)}x(b^{-1}a) \ee Then, from eqs.
(\ref{cocycle4}) and (\ref{3.132}) \be k_{a^{-1}b}=k_b \ee
Therefore $k_a=k_e$, and assuming $k_e=1$, it follows \be
(Cx)_a=x(a^{-1})=x(a)^{-1} \ee giving \be C_{a,b}=\delta_{ab,e}
\ee This shows also that \be C^T=C \label{3.144} \ee at least in
the cases of discrete and compact groups. The mapping
$C:\,\AAA\to\AAA$ is an involution of the algebra. In fact, by
defining \be x(a)^*=x(b)C_{b,a}=x(a^{-1})=x(a)^{-1}
\label{involution} \ee we have $(x(a)^*)^*=x(a)$, and
$x(b)^*x(a)^*=(x(a)x(b))^*$. From our general rule of integration
(see eq. (\ref{2.20})) we get
 \be
\int_{(x)}x(a)=C_{e,a}^{-1}=\delta_{e,a} \label{ATI} \ee
Therefore we are allowed to expand a function on the group ($\ket
f_a=f(a)$) as \be f(a)=\int_{(x)}x(a^{-1})\langle x|f\rangle
\label{2.43} \ee with $\langle x|f\rangle=\sum_{b\in G}
x(b)f(b)$. It is also possible to define a scalar product among
functions on the group. Defining,  $\bra f_a=\bar f(a)$, where
$\bar f(a)$ is the complex conjugated of $f(a)$, we put \be
\langle f |g\rangle=\int_{(x)} \langle f|Cx\rangle\langle
x|g\rangle= \int_{(x)}{\bar f}(x^*)g(x)=\sum_{a\in G} {\bar f}(a)
g(a) \ee It is important to stress that this definition depends
only on the algebraic properties of $\AAA(G)$ and not on the
specific representation chosen for this construction.

\subsection{What is the meaning of the algebraic integration?}

As we have said in the previous Section, the integration formula
we have obtained  is independent on the group representation we
started with. In fact, it is based only on the structure of right
and left multiplications, that is on the abstract algebraic
product. This independence on the representation suggests that in
some way we are "summing" over all the representations. To
understand this point, we will study in this Section vector
representations. To do that, let us introduce a label $\lambda$
for the vector representation we are actually using to define
$\AAA(G)$. Then a generic function on $\AAA(G)_\lambda$ \be \hat
f(\lambda)=\sum_{a\in G} f(a)x_\lambda(a) \label{F-transform} \ee
can be thought as the Fourier transform of the function $f:
G\to\complex$. Using the algebraic integration we can invert this
expression (see eq. (\ref{2.43})) \be f(a)=\int_{(x_\lambda)}\hat
f(\lambda) x_\lambda(a^{-1}) \label{inversion1} \ee But it is a
well known result of the  harmonic analysis over the groups, that
in many cases it is possible to invert the Fourier transform, by
an appropriate sum over the representations. This is true in
particular for finite and compact groups. Therefore the algebraic
integration should be the same thing as summing or integrating
over the labels $\lambda$ specifying the representation. In order
to show that this is the case, let us recall a few facts about
the Fourier transform over the groups.\cite{pont} First of all,
given the group $G$, one defines the set $\hat G$ of the
equivalence classes of the irreducible representations of $G$.
Then, at each point $\lambda$ in $\hat G$ we choose a unitary
representation $x_\lambda$ belonging to the class $\lambda$, and
define the Fourier transform of the function $f:G\to \complex$,
by the eq. (\ref{F-transform}). In the case of compact groups,
instead of the sum over the group element one has to integrate
over the group by means of the invariant Haar measure. For finite
groups, the inversion formula is given by \be f(a)=\frac
1{n_G}\sum_{\lambda\in\hat G} d_\lambda tr[\hat
f(\lambda)x_\lambda(a^{-1})] \label{3.3} \ee where $n_G$ is the
order of the group and $d_\lambda$ the dimension of the
representation $\lambda$. Therefore, we get the identification \be
\int_{(x)}\{\cdots\}=\frac 1{n_G}\sum_{\lambda\in\hat G}
d_\lambda tr[\{\cdots\}] \label{correspondence} \ee A more
interesting way of deriving this relation, is to take in
(\ref{F-transform}), $f(a)=\delta_{e,a}$, obtaining for its
Fourier transform, $\hat\delta=x_\lambda(e)=1_{\lambda}$, where
the last symbol means the identity in the representation
$\lambda$. By inserting this result into (\ref{3.3}) we get the
identity \be \delta_{e,a}=\frac 1{n_G}\sum_{\lambda\in\hat G}
d_\lambda tr[\hat x_\lambda(a^{-1})] \ee which, compared with eq.
(\ref{ATI}), gives (\ref{correspondence}). This  shows explicitly
that the algebraic integration for vector representations of $G$
is nothing but the sum over the representations of $G$.

An analogous relation is obtained in the case of compact groups.
This can also be obtained by a limiting procedure from  finite
groups, if we insert $ 1/{n_G}$, the volume of the group, in the
definition of the Fourier transform. That is one defines \be \hat
f(\lambda)=\frac 1 {n_G}\sum_{a\in G} f(a) x_\lambda(a) \ee from
which \be f(a)=\sum_{\lambda\in\hat G} d_\lambda tr[\hat
f(\lambda)x_\lambda(a^{-1})] \label{inversion2} \ee Then one can
go to the limit by substituting the  sum over the group elements
 with the  Haar measure
\be \hat f(\lambda)=\int_G d\mu(a) f(a) x_\lambda(a) \ee The
inversion formula (\ref{inversion2}) remains unchanged. We see
that in these cases the algebraic integration sums over the
elements of the space $\hat G$, and therefore it can be thought
as the dual of the sum over the group elements (or the Haar
integration for compact groups). By using the Fourier transform
(\ref{F-transform}) and its inversion (\ref{inversion1}), one can
easily establish the Plancherel fromula. In fact by multiplying
together two Fourier transforms, one gets \be \hat
f_1(\lambda)\hat f_2(\lambda)= \sum_{a\in G}\left(\sum_{b\in
G}f_1(b)f_2(b^{-1}a)\right)x_\lambda(a) \label{convolution} \ee
from which \be \int_{(x)}\hat f_1(\lambda)\hat
f_2(\lambda)x_\lambda(a^{-1})= \sum_{b\in G}f_1(b)f_2(b^{-1}a) \ee
and taking $a=e$ we obtain \be \int_{(x)}\hat f_1(\lambda)\hat
f_2(\lambda)= \sum_{b\in G}f_1(b)f_2(b^{-1})
\label{pre-Plancherel} \ee This formula can be further
specialized, by taking $f_2\equiv f$ and for $f_1$ the involuted
of $f$. That is \be \hat f^*(\lambda)=\sum_{a\in G} {\bar
f}(a)x_\lambda(a^{-1}) \ee where use has been made of eq.
(\ref{involution}). Then, from eq. (\ref{pre-Plancherel}) we get
the Plancherel formula \be \int_{(x)}\hat f^*(\lambda)\hat
f(\lambda)=\sum_{a\in G}{\bar f}(a)f(a) \label{Plancherel} \ee
Let us also notice that eq. (\ref{convolution})  says  that the
Fourier transform of the convolution of two functions on the
group is the product of the Fourier transforms.

We will consider now  projective representations.  In this case,
the product of two Fourier transforms is given by \be \hat
f_1(\lambda)\hat f_2(\lambda)= \sum_{a\in G} h(a) x_\lambda(a)
\label{product} \ee with \be h(a)=\sum_{b\in G}
f_1(b)f_2(b^{-1}a)\esp{i\alpha(b,b^{-1}a)} \label{convolution2}
\ee Therefore, for projective representations, the convolution
product is  deformed due to the presence of the phase factor.
However, the Plancherel formula still holds. In fact, since in \be
h(e)=\sum_{b\in G} f_1(b)f_2(b^{-1}) \ee using eq. (\ref{2.21}),
the phase factor disappears and the previous derivation from eq.
(\ref{pre-Plancherel}) to eq. (\ref{Plancherel}) is still valid.
Notice that eq. (\ref{product}) tells us that the Fourier
transform of the deformed convolution product of two functions on
the group, is equal to the product of the Fourier transforms.

\subsection{The case of abelian groups}

In this Section we consider the case of abelian groups, and we
compare the Fourier analysis made in our framework  with the more
conventional one made in terms of the characters. A fundamental
property of the abelian groups is that the set $\hat G$ of their
vector unitary irreducible representations (VUIR), is itself an
abelian group, the dual of $G$ (in the sense of Pontryagin\,\cite{pont}).
Since the VUIR's are one-dimensional, they are
given by the characters of the group. We will denote the
characters of $G$ by $\chi_\lambda(a)$, where $a\in G$, and
$\lambda$ denotes the representation of $G$. For what we said
before, the parameters $\lambda$ can be thought as the elements
of the dual group. The parameterization of the group element $a$
and of the representation label $\lambda$ are given in Table 1,
for the most important abelian groups and for their dual groups,
where we have used the notation $a=\vec a$ and $\lambda=\vec q$.

\begin{table}[t]

\begin{center}
 Table 1: {\it Parameterization of the abelian group $G$ and of its dual
$\hat G$,
 for $G=\real^D$, $Z^D$, $T^D$, $Z_n^D$}.
\end{center}
\vspace{0.4cm}
\begin{center}
\begin{tabular}{|c|c|c|c|c|}
\hline
& & & &\\
 & $G=R^D$& $G=Z^D$& $G=T^D$& $G=Z_N^D$\\
 &$\hat G= R^D$&$\hat G= T^D$&$\hat G= Z^D$&$\hat G= Z_N^D$\\
 &&&&\\
 \hline
 &&&&\\
 $\vec a$ & $-\infty\le a_i\le +\infty$ & $a_i=
 {\dd{\frac{2\pi m_i}L}}$ &
 $0\le a_i\le L $& $a_i= k_i,$\\
 && $m_i\in Z$ & & $0\le k_i\le n-1$\\
&&&&\\
\hline
 &&&&\\
 $\vec q$ & $-\infty\le q_i\le +\infty$ & $0\le q_i\le L $ &
 $q_i={\dd{\frac{2\pi m_i}L}}$&
  $q_i= {\dd{\frac{2\pi\ell_i}N}}$\\
 && &$m_i\in Z$  & $0\le \ell_i\le n-1$\\
&&&&\\
\hline
\end{tabular}
\end{center}
\end{table}

The characters are given by \be \chi_{\lambda}(a)\equiv\chi_{\vec
q}\;(\vec a)=\esp{-i\vec q\cdot\vec a} \ee and satisfy the
relation (here we  use the additive notation for the group
operation) \be \chi_\lambda(a+b)=\chi_\lambda(a)\chi_\lambda(b)
\label{4.3} \ee and the dual \be
\chi_{\lambda_1+\lambda_2}(a)=\chi_{\lambda_1}(a)\chi_{\lambda_2}(a)
\ee That is they define vector representations of the abelian
group $G$ and of its dual, $\hat G$. Also we can easily check
that the operators \be
 D_{\vec q} \chi_{\vec q}\;(\vec a)=-i\vec a \chi_{\vec q}\;(\vec a)
 \label{4.4}
\ee are derivations on the algebra (\ref{4.3}) of the characters
for any $G$ in Table 1.

We can use the characters to define the Fourier transform of the
function $f(g): G\to\complex$ \be \tilde f(\lambda)=\sum_{a\in G}
f(a)\chi_\lambda(a) \ee If we evaluate the Fourier transform of
the deformed convolution of eq. (\ref{convolution2}), we get \be
\tilde h(\lambda)=\sum_{a\in G} h(a)\chi_\lambda(a)= \sum_{a,b\in
G}f(a)\chi_\lambda(a) \esp{i\alpha (a,b)}g(b)\chi_\lambda(b) \ee
In the case of vector representations the Fourier transform of
the convolution is the product of the Fourier transforms. In the
case of projective representations, the result, using the
derivation introduced before, can be written in terms of the
Moyal product (we omit here the vector signs) \be \tilde
h(\lambda)=\tilde f(\lambda)\star\tilde
g(\lambda)=\esp{-i\alpha(D_{\lambda'},D_{\lambda''})} \tilde
f(\lambda')\tilde g(\lambda'')\Big|_{\lambda'=\lambda''= \lambda}
\ee Therefore, the Moyal product arises in a very natural way
from the projective group algebra. On the other hand,  we have
shown in the previous Section, that the use  of the Fourier
analysis in terms of the projective representations avoids the
Moyal product. The projective representations of abelian groups
allow a derivation on the algebra, analogous to the one in eq.
(\ref{4.4}), with very special features. In fact we check easily
that \be \vec D x_\lambda(\vec a)=-i\vec a x_\lambda(\vec a)
\label{derivation1} \ee is a derivation, and furthermore \be
\int_{(x_\lambda)} \vec D x_\lambda(\vec a)=0 \ee From this it
follows, by linearity, that the integral of  $\vec D$  applied to
any function on the algebra is zero \be \int_{(x_\lambda)} \vec D
\left(\sum_{a\in G}f(\vec a) x_\lambda(\vec a)\right)=0
\label{byparts} \ee This relation is very important because, as
we have shown in,\cite{algebra}  the automorphisms generated by
$\vec D$, that is $\exp(\vec\alpha\cdot\vec D)$, leave invariant
the integral. Notice that this derivation generalizes the
derivative with respect to the parameter $\vec q$, although this
has no meaning in the present case. In the case of nonabelian
groups, a derivation sharing the previous properties can be
defined only if there exists a mapping $\sigma: G\to \complex$,
such that \be \sigma(ab)=\sigma(a)+\sigma(b), ~~~~a,b\in G \ee
since in this case, defining \be Dx(a)=\sigma(a) x(a) \ee we get
\bea D(x(a)x(b))&=&\sigma(ab)
x(a)x(b)=(\sigma(a)+\sigma(b))x(a)x(b)\nn\\&=&
(Dx(a))x(b)+x(a)(Dx(b)) \eea

Having defined derivations and integrals one has all the elements
for the harmonic analysis on the projective representations of an
abelian group.

Let us start considering $G=R^D$. In the case of vector
representations we have \be x_{\vec q}\;(\vec a)=\esp{-i\vec
q\cdot\vec a} \label{a0} \ee with $\vec a\in G$, and $\vec
q\in\hat G=R^D$ labels the representation. The Fourier transform
is \be \hat f(\vec q)=\int d^D\vec a f(\vec a)\esp{-i\vec
q\cdot\vec a} \label{a1} \ee Here the Haar measure for $G$
coincides with the ordinary Lebesgue measure. Also, since $\hat
G=R^D$, we can invert the Fourier transform by using the Haar
measure on the dual group, that is, again the Lebesgue measure.
In the projective case, eq. (\ref{a0}) still holds true, if we
assume $\vec q$
 as
a vector operator satisfying the commutation relations \be
[q_i,q_j]=i\eta_{ij} \label{CR} \ee with $\eta_{ij}$ numbers
which can be related to the cocycle, by using the
Baker-Campbell-Hausdorff formula \be \esp{-i\vec q\cdot\vec
a}\esp{-i\vec q\cdot\vec b}= \esp{-i\eta_{ij}a_ib_j/2}\esp{-i\vec
q\cdot(\vec a+\vec b)} \ee giving \be \alpha(\vec a,\vec
b)=-\frac 1 2 \eta_{ij} a_i b_j \ee The inversion of the Fourier
transform can now be obtained by our formulation of the algebraic
integration in the form \be f(\vec a)=\int_{(\vec q)}\hat f(\vec
q)x_{\vec q}\;(-\vec a) \ee where the dependence on the
representation is expressed in terms of $\vec q$, thought now
they are not coordinates on $\hat G$. We recall that in this
case, eq. (\ref{ATI}) gives \be \int_{(\vec q)}x_{\vec q}\;(\vec
a)=\delta^D(\vec a) \label{a2} \ee Therefore, the relation
between the algebraic integration  and the Lebesgue integral in
$\hat G$, in the vector case is \be \int_{(\vec
q)}=\int\frac{d^D\vec q}{(2\pi)^D} \ee In the projective case
the  right hand side of this relation has no meaning, whereas the
left hand side is still well defined. Also, we cannot maintain
the interpretation of  the $q_i$'s as coordinates on the dual
space $\hat G$. However, we can define elements of $\AAA(G)$
having the properties of the $q_i$'s (in particular satisfying
eq. (\ref{CR})), by using the Fourier analysis. That is we define
\be q_i=\int d^D\vec a\left(-i\frac{\partial}{\partial
a_i}\delta^D(\vec a)\right)x_{\vec q}\;(\vec a) \label{q-def} \ee
which is  an element of $\AAA(G)$ obtained by Fourier
transforming a distribution  over $G$, which is a honestly
defined space. From this definition we can easily evaluate the
product \be q_i x_{\vec q}\;(\vec a)=\int d^D\vec
b\left(-i\frac{\partial}{\partial b_i}\delta^D(\vec
b)\right)x_{\vec q}\;(\vec b)x_{\vec q}\;(\vec a) \ee Using the
algebra  and integrating by parts, one gets the result \be
q_ix_{\vec q}\;(\vec a)=i\nabla_i x_{\vec q}\;(\vec a)
\label{5.28} \ee where \be \nabla_i=\frac{\partial}{\partial a_i}
+i\alpha_{ij} a_j \ee where $\alpha_{ij}=\alpha({\vec
e}_{(i)},{\vec e}_{(j)})$, with ${\vec e}_{(i)}$ an orthonormal
basis in $\real^D$. In a completely analogous way one finds \be
x_{\vec q}\;(\vec a)q_i=i{\overline \nabla}_i x_{\vec q}\;(\vec a)
\ee where \be {\overline\nabla}_i=\frac{\partial}{\partial a_i}
-i\alpha_{ij} a_j \ee Then, we  evaluate the  commutator \be
[q_i,\hat f(\vec q)]=\int d^D\vec a
\left[-i\left({\overline\nabla}_i-\nabla_i\right)f(\vec
a)\right]x_{\vec q}(\vec a) \ee where we have done an integration
by parts. We get \be [q_i,\hat f(\vec
q)]=-2i\alpha_{ij}D_{q_j}\hat f(\vec q) \label{5.30} \ee where
$D_{q_j}$ is the derivation (\ref{derivation1}), with $q_j$ a
reminder for the direction along wich the derivation acts upon.
In particular, from \be D_{q_j}q_i=\int d^D\vec
a\left(-i\frac{\partial}{\partial a_i}\delta^D(\vec
a)\right)(-ia_j)x_{\vec q}(\vec a)=\delta_{ij} \ee we get \be
[q_i,q_j]=-2i\alpha_{ij} \ee in agreement with eq. (\ref{CR}),
after the identification $\alpha_{ij}=-\eta_{ij}/2$.

The automorphisms induced by the derivations (\ref{derivation1})
are easily evaluated \be S(\vec \alpha)x_{\vec q}\;(\vec
a)=\esp{\vec\alpha\cdot D_{\vec q}} x_{\vec q}\;(\vec
a)=\esp{-i\vec\alpha\cdot\vec a}x_{\vec q}\;(\vec a)=x_{\vec
q+\vec\alpha}(\vec a) \label{5.231} \ee where the last equality
follows  from \be \int d^D\vec a\left(-i\frac{\partial}{\partial
a_i}\delta^D(\vec a)\right)\esp{\vec\alpha\cdot D_{\vec
q}}x_{\vec q}\;(\vec a)= q_i+\alpha_i \ee Meaning that in the
vector case,  $S(\vec \alpha)$ induces translations in $\hat G$.
Since $D_{\vec q}$ satisfies the eq. (\ref{byparts}), it follows
from Section 2.4 (see also\,\cite{algebra}) that the automorphism
$S(\vec\alpha)$ leaves invariant the algebraic integration
measure \be \int_{(\vec q)}=\int_{(\vec q+\vec\alpha)}
\label{invariance1} \ee This shows that it is possible to
construct a calculus completely analogous to the one that we have
on $\hat G$ in the vector case, just using the Fourier analysis
following by the algebraic definition of the integral. We can
push this analysis a little bit further by looking  at the
following expression \be \int_{(\vec q)}\hat f(\vec
q)\;q_i\;x_{\vec q}\;(-\vec a)=-i \left(\frac{\partial}{\partial
a_i}+i \alpha_{ij}a_j\right)f(\vec a) \label{magnetic} \ee where
we have used  eq. (\ref{5.28}). In the case $D=2$ this equation
has a physical interpretation in terms of a  particle of charge
$e$, in a constant magnetic field $B$. In fact, the commutators
among canonical momenta are \be [\pi_i,\pi_j]=ieB\epsilon_{ij}
\ee where $\epsilon_{ij}$ is the 2-dimensional Ricci tensor.
Therefore, identifying $\pi_i$ with $q_i$, we get
$\alpha_{ij}=-eB\epsilon_{ij}/2$. The corresponding vector
potential is given by \be A_i(\vec a)=-\frac 1
2\epsilon_{ij}Ba_j=\frac 1 {e}\alpha_{ij}a_j \ee Then, eq.
(\ref{magnetic}) tells us that the operation $\hat f(\vec q)\to
\hat f(\vec q)q_i$, corresponds to  take the covariant derivative
\be -i\frac{\partial}{\partial a_i}+eA_i(\vec a) \ee of the
inverse Fourier transform of $\hat f(\vec q)$.  An interesting
remark is that a translation in $\vec q$ generated by
$\exp(\vec\alpha\cdot \vec D)$, gives rise to a phase
transformation on  $f(\vec a)$. First of all, by using the
invariance of the integration measure we can check that \be \hat
f(\vec q+\vec\alpha)=\esp{\vec\alpha\cdot\vec D} \hat f(\vec q)
\label{equality} \ee In fact \be \int_{(\vec q)}\hat f(\vec
q+\vec\alpha)x_{\vec q}\;(-\vec a)= \int_{(\vec
q-\vec\alpha)}\hat f(\vec q)x_{\vec q-\vec\alpha}\;(-\vec a)=
\esp{-i\vec\alpha\cdot\vec a}f(\vec a) \ee Then, we have \be
\int_{(\vec q)} \left(\esp{\vec\alpha\cdot\vec D}\hat f(\vec
q)\right)x_{\vec q}\;(-\vec a)= \int_{(\vec q)} \hat f(\vec
q)\left(\esp{-\vec\alpha\cdot\vec D} x_{\vec q}(-\vec a)\right)=
\esp{-i\vec \alpha\cdot\vec a}f(\vec a) \ee where we have made
use of eq. (\ref{5.231}). This proves  eq. (\ref{equality}), and
at the same time our assertion.  From eq. (\ref{magnetic}), this
is equivalent to a  gauge transformation on the gauge potential
${\cal A}_i=\alpha_{ij}a_j$, ${\cal A}_i\to{\cal A}_i-
\partial_i\Lambda$, with $\Lambda=\vec \alpha\cdot\vec a$.
Therefore, we see here explicitly the content of a projective
representation in the basis of the functions on the group. One
starts assigning the two-form $\alpha_{ij}$. Given that, one
makes a choice for the vector potential. For instance in the
previous analysis we have chosen $\alpha_{ij}a_j$. Any possible
projective representation corresponds to a different choice of
the gauge.  In the dual Fourier basis  this corresponds to assign
a fixed set of operators $q_i$, with commutation relations
determined by a two-form. All the possible projective
representations are obtained by translating the operators
$q_i$'s. Of course, this is equivalent to say that the projective
representations are the central extension of the vector ones, and
that they are determined by the cocycles. But the previous
analysis shows that the projective representations generate
noncommutative spaces, and that the algebraic integration,
allowing us to define a Fourier analysis, gives the possibility
of establishing  calculus rules over these spaces.

Consider now the case $G=Z^D$. Let  us introduce an orthonormal
basis on the square lattice defined by $Z^D$, ${\vec e}_{(i)}$,
$i=1,\cdots,D$. Then, any element of the algebra can be
reconstructed in terms of a product of the elements \be
U_i=x({\vec e}_{(i)}) \label{definition} \ee corresponding to a
translation along the  direction $i$ by one lattice site. In
general we will have \be x(\vec m)=\esp{i\theta(\vec
m)}U_1^{m_1}\cdots U_D^{m_D},~~~~ \vec m=\sum_i m_i{\vec e}_{(i)}
\label{4.42} \ee with $\theta$ a calculable phase factor. The
quantities $U_i$ play the same role of $\vec q$ of the previous
example. The Fourier transform is defined by \be \hat f(\vec
U)=\sum_{\vec m\in Z^D} f(\vec m) x_{\vec U}(\vec m) \ee where
the dependence on the representation is expressed in terms of
$\vec U$, denoting the collections of the $U_i$'s. The inverse
Fourier transform is defined by \be f(\vec m)=\int_{\vec U} \hat
f(\vec U)x_{\vec U}(-\vec m) \ee where the integration rule is \be
\int_{(\vec U)} x_{\vec U}(\vec m)=\delta_{\vec m,\vec 0} \ee
Therefore, the Fourier transform of $U_i$ is simply $\delta_{\vec
m, \vec e_{(i)}}$. The algebraic integration for the vector case
is \be \int_{(\vec U)}\to \int_0^{L} \frac{d^D\vec q}{L^D} \ee
Since the set $\vec U$ is within the generators of the algebra,
to establish the rules of the calculus is a very simple matter.
Eq. (\ref{definition}) is the definition of the set $\vec U$,
analogous to eq. (\ref{q-def}). In place of eq. (\ref{5.30}) we
get \be U_i\hat f(\vec U)U_i^{-1}= \esp{-2\alpha_{ij}D_j}\hat
f(\vec U) \ee Here $D_j$ is the $j$-th  component of the
derivation $\vec D$ which acts upon $U_i$ as \be
D_iU_j=-i\delta_{ij}U_j \ee By choosing $\hat f(\vec U)=U_k$ we
have \be U_iU_kU_i^{-1}U_k^{-1}=\esp{2 i\alpha_{ik}} \ee which is
the analogue of  the commutator among the $q_i$'s. The
automorphisms generated by $\vec D$ are \be S(\vec\phi)x_{\vec
U}(\vec m)=\esp{\vec\phi\cdot\vec D}x_{\vec U}(\vec m)=
\esp{-i\vec\phi\cdot\vec m}x_{\vec U}(\vec m) \ee From which we
see that \be U_i\to S(\vec\phi)U_i=\esp{-i\phi_i}U_i \ee This
transformation corresponds to  a trivial cocycle.  As in the case
$G=\real^D$ it gives rise to a phase transformation on the group
functions \be \int_{(\vec U)}\left(\esp{\vec\alpha\cdot\vec
D}\hat f(\vec U)\right)x_{\vec U}(-\vec m)= \int_{\vec U)}\hat
f(\vec U)\left(\esp{-\vec\alpha\cdot\vec D}x_{\vec U} (-\vec
m)\right)= \esp{i\vec\phi\cdot\vec m} f(\vec m) \ee Of course,
all these relations could be obtained formally, by putting
$U_i=\exp(-iq_i)$, with $q_i$ defined as in the case $G=R^D$.

Finally, in the case $G=Z_n^D$, the situation is very much alike
$Z^D$, that is the algebra can be reconstructed in terms of a
product of elements \be U_i=x({\vec e}_{\;(i)}) \ee satisfying \be
U_i^n=1 \ee Therefore we will not repeat the previous analysis
but we will consider only the case $D=2$, where $U_1$ and $U_2$
can be expressed as\,\cite{hoppe} \be
(U_1)_{a,b}=\delta_{a,b-1}+\delta_{a,n}\delta_{b,1},~~~~
(U_2)_{a,b}=\esp{\frac {2\pi i}
n(a-1)}\delta_{a,b},~~~~a,b=1,\cdots,n \ee The elements of the
algebra are reconstructed as \be x_{\vec U}(\vec
m)=\esp{i\frac{\pi} n m_1m_2}U_1^{m_1}U_2^{m_2} \ee The cocycle
is now \be \alpha(\vec m_1,\vec m_2)=-\frac{2\pi} n\epsilon_{ij}
m_{1i}m_{2j} \ee In this case we can compare the algebraic
integration rule \be \int_{\vec U} x_{\vec U}(\vec
m)=\delta_{\vec m,\vec 0} \ee with \be Tr[x_{\vec U}(\vec
m)]=n\delta_{\vec m,\vec 0} \ee A generic element of the algebra
is a $n\times n$ matrix \be
A=\sum_{m_1,m_2=0}^{n-1}c_{m_1m_2}x_{\vec U}(\vec m) \ee and
therefore \be \int_{\vec U} A=\frac 1 n Tr[A] \ee In Section 3 we
have shown that the algebraic integration over the algebra of the
$n\times n$ matrices $\AAA_n$ is given by \be \int_{\AAA_n}
A=Tr[A] \ee implying \be \int_{\vec U}A=\frac 1 n \int_{\AAA_n}A
\ee

\subsection{The example of the algebra on the circle}

A particular example of a group algebra is the algebra on the
circle defined by \be z^nz^m=z^{n+m},~~~-\infty\le n,m\le +\infty
\ee with $z$ restricted to the unit circle. \be z^*=z^{-1} \ee
This is a group algebra over $\Zint$. Defining the ket \be \ket
z=\left(\matrix{\cdot\cr z^{-i}\cr\cdot\cr 1\cr z\cr\cdot\cr
z^i\cr\cdot\cr}\right) \ee the $R_i$ and $L_i$ matrices are given
by \be (R_i)_{jk}=\delta_{i+j,k},~~~(L_i)_{jk}=\delta_{i+k,j} \ee
and from our previous construction, the  matrix $C$ is given by
\be (C)_{ij}=(C^{-1})_{ij}=\delta_{i+j,0} \ee or. more explicitly
by \be C=\left(\matrix{\cdot & \cdot & \cdot & \cdot & \cdot \cr
                \cdot &   0   &    0  &   1   & \cdot \cr
                \cdot &   0   &    1  &   0   & \cdot \cr
                \cdot &   1   &    0  &   0   & \cdot \cr
                \cdot & \cdot & \cdot & \cdot & \cdot \cr }\right)
\ee showing that \be C:~~~z^i\to z^{-i} \ee In fact \be
(C^{-1}L_i C)_{lp}=\delta_{l,-m}\delta_{i+n,m}\delta_{n,-p}=
\delta_{i-p,-l}=\delta_{i+l,p}=(R_i)_{lp} \ee In this case the
$C$ matrix is nothing but the
 complex conjugation ($z\to z^*=z^{-1}$).
 The completeness relation reads now
\be \int_{(z)}z^iz^{-j}=\delta_{ij} \ee from which \be
\int_{(z)}z^k=\delta_{k0} \ee
 Our algebraic definition of integral can be interpreted as
an integral along a circle $C$ around the origin. In fact we have
\be \int_{(z)}=\frac{1}{2\pi i}\int_C\frac{dz}{z} \ee
\section{Associative Non Self-Conjugated Algebras: the $q$-oscillator}


A generalization of the bosonic oscillator is the $q$-bosonic
oscillator.\cite{biedenharn} We will use the definition given
in\,\cite{baulieu}
\be b\bar b-q\bar b b=1 \ee with $q$ real and
positive. We assume as elements of the algebra ${\cal A}$, the
quantities \be x_i=\frac {z^i}{\sqrt{i_q!}} \ee where $z$ is a
complex number, \be i_q=\frac {q^i-1}{q-1} \ee and \be
i_q!=i_q(i-1)_q\cdots 1 \ee The structure constants are \be
f_{ijk}=\delta_{i+j,k}\sqrt{\frac{k_q!}{i_q!j_q!}} \ee and
therefore \be
(R_i)_{jk}=\delta_{i+j,k}\sqrt{\frac{k_q!}{i_q!j_q!}},~~~~
(L_i)_{jk}=\delta_{i+k,j}\sqrt{\frac{j_q!}{i_q!k_q!}} \ee In
particular \be (R_1)_{jk}=\delta_{j+1,k}\sqrt{k_q}, ~~~~~
(L_1)_{jk}=\delta_{j-1,k}\sqrt{(k+1)_q} \ee We see that $R_1$ and
$L_1$ satisfy the $q$-bosonic algebra \be R_1L_1-qL_1 R_1=1 \ee
This equation shows that the right- and left-representations are
not equivalent  for $q\not =-1$ (the Fermi oscillator case).
Therefore no $C$ matrix exists and, according to our rules of
Section 2.2, we require\be
\int_{(z,z^*)_q}\frac{z^i{z^*}^j}{i_q!j_q!}=\delta_{ij} \ee In
the case $q=1$ the integration coincides with the standard
integration over complex numbers as used for coherent
states\,\cite{integrale} \be \int_{(z,z^*)_1}=\int\frac{dz\,dz^*}{2\pi i}
\esp{-|z|^2}\ee

The integration for the $q$-oscillator can be expressed in terms
of the so called $q$-integral (see ref.\cite{koorwinder}), by
using the representation of $n_q!$ as a $q$-integral \be
n_q!=\int_0^{1/(1-q)}d_qt\; e_{1/q}^{-qt}\; t^n \ee where the
$q$-exponential is defined by \be
e_{q}^{t}=\sum_{n=0}^\infty\frac{z^n}{n_q!} \ee and the
$q$-integral through \be \int_0^a d_qt
f(t)=a(1-q)\sum_{n=0}^\infty f(aq^n)q^n \ee Then the two
integrations are related by ($z=|z|\exp(i\phi)$)
 \be
 \int_{(z,z^*)_q}=\int\frac{d\phi}{2\pi}\int d_q(|z|^2)\;
 e_{1/q}^{-q|z|^2}
 \ee
 The Jackson integral is the inverse of the $q$-derivative
\be (D_qf)(x)=\frac{f(x)-f(qx)}{(1-q)x} \ee In fact, for \be
F(a)=\int_0^a d_qt f(t)=a(1-q)\sum_{n=0}^\infty f(aq^n)q^n \ee
one has \be (D_qF)(a)=f(a) \ee as can be checked using \be
F(qa)=qa(1-q)\sum_{n=0}^\infty f(aq^{n+1})q^n
=a(1-q)\sum_{n=1}^\infty f(aq^n)q^n=F(a)-a(1-q)f(a) \ee

The limit $q\to 1$ defines the rules for the integration in the
case of the normal bosonic oscillator.
\section{Non-Associative  Self-Conjugated Algebras:
the octonions}


We will discuss  now how to integrate over the  non-associative
algebra of octonions (see\,\cite{ottonioni}). This algebra (said
also a Cayley algebra) is defined in terms of the multiplication
table of its seven imaginary units $e_A$ \be e_Ae_B=-\delta_{AB}+
a_{ABC}e_C,~~~A,B,C=1,\cdots,7 \ee where $a_{ABC}$ is completely
antisymmetric and equal to +1 for $(ABC)=(1,2,3)$, (2,4,6),
(4,3,5), (3,6,7), (6,5,1), (5,7,2) and (7,1,4). The automorphism
group of the algebra is $G_2$. We define  the split basis as \bea
u_0=\frac 1 2 (1+ie_7),&&u_0^*=\frac 1 2 (1-ie_7)\nn\\
u_i=\frac 1 2 (e_i+ie_{i+3}),&&u_i^*=\frac 1 2 (e_i-ie_{i+3}) \eea
where $i=1,2,3$. In this basis  the  multiplication rules are
given in Table 2 and automorphism group is $SU(3)$.
\begin{center}\bigskip
\vspace{0.2cm} Table 2: {\it Multiplication table for the
octonionic algebra.}
\end{center}
\vspace{0.4cm}
\begin{center}
\begin{tabular}{|c|c|c|c|c|}
\hline
 &~~~$u_0$~~~  &~~~ $u_0^*$~~~ &$u_j$& $u_j^*$ \\
\hline
$u_0$ &$u_0$ & $0$ & $u_j$ &$0$ \\
\hline
$u_0^*$ &$0$ &$u_0^* $&$0$  &$u_j^*$ \\
\hline
$u_i$ & $0$ &  $u_i$ &~ $\epsilon_{ijk}u_k^*$~ & $-\delta_{ij}u_0$\\
\hline
$u_i^*$& $u_i^*$ &$0$ & $-\delta_{ij}u_0^*$ &~ $\epsilon_{ijk} u_k$~\\
\hline
\end{tabular}
\end{center}
\vspace{0.4cm}   The non-associativity can be checked by taking,
for instance, \be u_i(u_j u^*_k)=u_i(-\delta_{jk} u_0)=0 \ee and
comparing with \be (u_i u_j)u^*_k=\epsilon_{ijm}u_m^*
u^*_k=-\epsilon_{ijk}\epsilon_{kmn}u_n \ee From the ket \be \ket
u=\left(\matrix{u_0\cr u_0^*\cr u_i\cr u_i^*\cr}\right) \ee one
can easily evaluate the matrices $R$ and $L$ corresponding to
right and left multiplication.
We will not give here the explicit expressions, but one can
easily see some properties. For instance, one can evaluate the
anticommutator $[R_i,R^*_j]_+$, by using the following relation
\be [R_i,R^*_j]_+\ket u=R_i\ket u u_j^*+R_j^*\ket u u_i= (\ket u
u_i)u_j^*+(\ket u u_j^*) u_i \ee The algebra of the
anticommutators of $R_i,R_i^*$ turns out to be the algebra of
three Fermi oscillators (apart from the sign) \be
[R_i,R^*_j]_+=-\delta_{ij},~~~ [R_i,R_j]_+=0,~~~[R_i^*,R^*_j]_+=0
\ee The matrices $R_0$ and $R_0^*$ define orthogonal projectors
\be R_0^2=R_0,~~~ (R_0^*)^2=R_0^*,R_0R_0^*=R_0^*R_0=0 \ee Further
properties are \be R_0+R_0^*=1 \ee and \be R_i^*=-R_i^T \ee
Similar properties hold for the left multiplication matrices.
This algebra is self-conjugated with the $C$ matrix given by\be
C=\left(\matrix{1 & 0 & 0 & 0\cr
        0 & 1 & 0 & 0\cr
        0 & 0 & 0 & -1_3\cr
        0 & 0 & -1_3 &0\cr}\right)
\ee where $1_3$ is the $3\times 3$ identity matrix. We have
$C^T=C$. The integration rules are easily obtained by looking at
the following external product \bea C\ket u\bra
u&=&\left(\matrix{u_0\cr u_0^*\cr -u_i^*\cr -u_i}\right)
\left(u_0,u_0^*,u_j,u_j^*\right)\nn\\&=& \left(\matrix{ u_0 & 0 &
u_j & 0\cr
                0  & u_0^* & 0 & u_j^*\cr
                -u_i^*  & 0 & \delta_{ij} u_0^* & -\epsilon_{ijk} u_k\cr
                0 & -u_i & -\epsilon_{ijk} u_k^* &
                     \delta_{ij} u_0\cr}
\right) \eea Therefore we get \be
\int_{(u)}\,u_0=\int_{(u)}\,u_0^*=1,~~~~
\int_{(u)}\,u_i=\int_{(u)}\,u_i^*=0 \ee




\section*{References}
\addcontentsline{toc}{section}{\numberline{}References}

\end{document}